\documentclass{amsart}
\calclayout

\usepackage{siunitx, commath}
\usepackage{xcolor,graphicx}
\usepackage[linesnumbered,algoruled,boxed,lined]{algorithm2e}
\usepackage{bm}
\makeatletter
\@tfor\next:=abcdefghijklmnopqrstuvwxyzABCDEFGHIJKLMNOPQRSTUVWXYZ\do{%
  \def\command@factory#1{%
    \expandafter\def\csname V#1\endcsname{\mathbf{#1}}
  }
 \expandafter\command@factory\next
}
\def\greekvectors#1{%
 \@for\next:=#1\do{%
    \def\X##1;{%
     \expandafter\def\csname V##1\endcsname{\bm{\csname##1\endcsname}}
     }
   \expandafter\X\next;
  }
}
\greekvectors{alpha,beta,gamma,delta, epsilon,varepsilon,varphi,iota,lambda,nu,
eta,varsigma,sigma,phi,psi,Delta,Sigma,mu,Phi}

\newcommand{\pb}[1]{\left[ #1 \right]}
\newcommand{\uinc}{u^{\mathrm{inc}}}
\newcommand{\usc}{u^{\mathrm{s}}}
\newcommand{\kin}{k_1}
\newcommand{\kout}{k_0}
\newcommand{\overbar}[1]{\mkern 1.5mu\overline{\mkern-1.5mu#1\mkern-1.5mu}\mkern 1.5mu}

\begin{document}

\title{Efficient computational design and optimization of dielectric metamaterial devices}

\author{Boaz Blankrot and Clemens Heitzinger}

\address{Institute of Analysis and Scientific Computing, TU Wien, A-1040 Vienna, Austria}

\email{boaz.blankrot@tuwien.ac.at} 

\begin{abstract}
Dielectric structures composed of many inclusions that manipulate light in ways the bulk materials cannot are commonly seen in the field of metamaterials.
In these structures, each inclusion depends on a set of parameters such as location and orientation, which are difficult to ascertain.
We propose and implement an optimization-based approach for designing such metamaterials in two dimensions by using a fast boundary element method and a multiple-scattering solver for a given set of parameters.
This approach provides the backbone of an automated process for the design and analysis of metamaterials that does not rely on analytical approximations.
We demonstrate the validity of our approach with simulations that converge to optimal parameter values and result in substantially better performance.
\end{abstract}

\maketitle

\section{Introduction}

Over the past decade, interest in metamaterials and specifically in dielectric metamaterials has grown considerably.
Initially, much of the research efforts were focused on exotic applications attained by negative-index metallic-based metamaterials such as cloaking \cite{ar:cloaking} and perfect lenses \cite{ar:pendry}. In recent years, much of the focus has shifted to dielectric metamaterials which are better-behaved with respect to power dissipation \cite{ar:jahani} and are easily fabricated \cite{ar:yang2014, ar:moitra2015}.
One prominent example of dielectric nanostructures is dielectric photonic crystals, which have been intensively investigated over the past thirty years~\cite{ar:yablonovitch1994, bk:joannopoulos2011}. Photonic crystals are composed of a one- to three-dimensional periodic array of nanostructures, in which a small number of cells may be altered or defective. This structure is designed to allow, alter, or prevent the propagation of light for a selected range of wavelengths. These nanostructures can be e.g.~round holes~\cite{ar:yablonovite} or contain a complex network of nano-engineered rods~\cite{ar:turner2013, ar:taverne2016}.
Thanks to their ability to control light flow, photonic crystals have promising applications in the developing field of optical computing. Replacing electronic components in integrated circuits with their photonic crystal counterparts will reduce the size and latencies of computer processors, while substantially increasing power efficiency \cite{ar:yablonovitch2002photonic, ar:Cuesta-Soto:04}.

Dielectric metalenses are another class of recently popular all-dielectric metamaterials~\cite{ar:metalens_review}. These metalenses allow manipulation of light for many practical applications, including chirality imaging~\cite{ar:khorasaninejad2016chiral}, imaging with reduced aberrations~\cite{ar:arbabi2016miniature}, and optical fiber coupling~\cite{ar:metalens_fiber}. Metalenses are typically comprised of numerous sub-wavelength building blocks arranged on a substrate.
The properties of these building blocks dictate which effect the overall metalens has on light passing through it.
There are many degrees of freedom in designing metalenses; the size, shape, rotation, and material of each individual building block can be adjusted arbitrarily, yielding a large variety of possible metalenses to meet different objectives~\cite{ar:metasurfaces2014}.
Nevertheless, this freedom creates a large search space in choosing these parameters, which may number in the thousands.

Optimization methods have been employed in the past for designing optical structures. For example, dielectric antireflective layers with piecewise constant permittivity were designed with a gradient descent algorithm~\cite{ar:dobson1993}.
Optimization has been combined with the Finite-Difference Time-Domain method for optimizing mode confinement in photonic crystal cavities~\cite{ar:bauer2008}, however from a computational perspective, this approach may suffer from the need to re-solve the entire problem when changes are made to the parameters.
Optimization of photonic crystal devices with circular inclusions was performed in~\cite{ar:cao2014} by means of transformation optics.
Spatial optimization of circular dielectric rods in the radio-frequency regime was performed in \cite{ar:seliger2006, ar:bertsimas2007} using a finite differences discretization and both gradient-based and gradient-free algorithms.
Shape and topology optimization for optical structures is fairly established, both in the periodic~\cite{ar:bao1998modeling, ar:yablonovitch2005}
and non-periodic cases~\cite{ar:jesselu2013}. In \cite{ar:miller2013adjoint}, shape optimization was accelerated with the adjoint-state method to reduce the number of necessary field computations.

We propose a specialized optimization-based method for analyzing and designing metamaterials in an automated fashion.
The class of problems we consider consists of metamaterials with a large number of inclusions, which may be circular, but the number of non-circular prototype inclusions is small relative to the number of inclusions.
We say that differently oriented inclusions of the same shape and material have the same prototype.
This approach utilizes a multipole expansion and a fast multiple-scattering method~\cite{ar:gumerov2007, ar:lai14} to solve the underlying electromagnetic problem, and a gradient-based algorithm for the optimization. 
Our approach is most appropriate for optimizing radii in case of circular inclusions, and for optimizing rotation angles in case of general inclusions, which corresponds to the design of many photonic crystals and metamaterials.
In the context of large-scale, aperiodic metamaterials in the class above, this type of efficient automation of the design process for specified optical properties has not been previously proposed.

The remainder of the paper is organized as follows.
Section~\ref{sec:description} gives the problem description and method overview.
The mathematical formulation used for calculating the fields scattered by a collection of inclusions is presented in detail in Section~\ref{sec:math}.
Section~\ref{sec:optimization} presents our optimization framework for the automated design of metamaterials, which is given as pseudocode in Algorithm~\ref{algo:1}.
Numerical results of both rotation angle and radius optimization are shown in Section~\ref{sec:results}, as well as a time complexity analysis of our approach. The results of this work are summarized in Section~\ref{sec:conc}.

\section{Problem and method description}\label{sec:description}
The problems solved in this work consist of a layout of smooth inclusions which may be circular, and an objective function that models a desired electric field distribution  at a set of points of interest. Our goal is to simultaneously optimize the radius of each circular inclusion and the rotation angle of each non-circular one to fit some desired behavior.

In this work, problems are restricted to time-harmonic incident fields scattering off a collection of two-dimensional inclusions in free space, where the variation $\exp(-i \omega t)$ is assumed and suppressed. We restrict this treatment to TM waves with respect to $z$, but the TE formulation is readily available with small modifications. We assume $M$ inclusion surfaces $\Omega_m$ with smooth boundaries $\partial \Omega_m$, in which the wavenumber $k_m = \omega \sqrt{\mu_0 \varepsilon_m}$ is real and constant, and $\Omega_0$ denotes the open free-space domain.
Hence the $\hat{\Vz}$ component of the electric field is the solution of the Helmholtz equation
\begin{align}
\nabla^2 u + k_m^2 u = 0, \quad
u=
\begin{cases}
\uinc + \usc &\mbox{in } \Omega_{0}, \\
\usc &\mbox{in } \Omega_{m \neq 0},
\end{cases}
\end{align}
where $\uinc$ is the given incident field, $\usc$ is the scattered field, and the jump in both $u$ and the normal derivative $\partial u / \partial n$ is zero across all boundaries, corresponding to continuity of the tangential electric field and the normal magnetic flux density. In addition, the scattered field must satisfy the Sommerfeld radiation condition in $\Omega_0$, but this is automatically satisfied due to the integral equation method used here.
We assume an objective function that depends on the electric field intensity at multiple points, of the form
\begin{align}
\label{eq:fobj}
f_{\mathrm{obj}} := \sum_{i = 1}^{I} |u(\Vr_i)|^2
,
\end{align}
where other functions of the intensity can be optimized via the chain rule.

We give an overview of our method. First, we use a boundary integral equation~\cite{bk:kress2013} to discretize each prototype inclusion once and transform it to a compressed cylindrical harmonics representation. It is straightforward to rotate and move this representation. 
We then apply a multiple-scattering approach~\cite{ar:lax1951, ar:gumerov2007} on these representations in order to describe the electromagnetic interactions between the inclusions. Once we solve the arising multiple-scattering problem with the Fast Multipole Method (FMM)~\cite{ar:coifman1993}, we can easily compute the electromagnetic field at any point.
This combination of boundary integral equation and multiple-scattering methods was applied to thin strips~\cite{ar:chew1991}, three-dimensional scattering~\cite{ar:martin2003, ar:greengard2013}, and two-dimensional multi-layered structures~\cite{ar:lai14}.
The computational complexity of this step is sufficiently low for employing optimization methods that require many solutions, as the ability to quickly compute the field at any collection of points makes it simple to define and compute an objective function for minimizing and/or maximizing the field intensity at multiple points. 
The integral equation approach naturally begets gradient-based optimization, which converges to a locally optimal set of parameters and yields an exact result in each step. We speed up the gradient-based optimization with the adjoint-state method~\cite{ar:chavent1974identification} (see also~\cite{ar:plessix2006adjoint}) which significantly decreases the optimization run time.

\section{Scattering formulation}
\label{sec:math}
In this section, we describe the mathematical background used to calculate the field scattered by a collection of inclusions at any point. First, we handle the case of a single inclusion, then we provide the formulation used for multiple scattering, and lastly we apply FMM to accelerate the solution process. 
The mathematical development of the single inclusion and multiple-scattering formulations follow that of~\cite{ar:lai14}, and is repeated here for ease of reading.

\subsection{Single inclusion formulation}
\label{sec:method_single}
First we apply Nystr{\"o}m discretization to a single prototype inclusion and transform its representation from that of boundary potential densities to cylindrical harmonics.
There are three motivations for this transformation. For smooth inclusions, the number of discretization nodes is dramatically larger than the number of cylindrical harmonics, which allows us to precompute the transformation for each inclusion shape once and only deal with the cylindrical harmonics representation without increasing the error in the electric field. This reduces the computational cost of the solution to a multiple-scattering problem by several orders of magnitude, and is particularly helpful when multiple iterations of a scattering problem are required for optimization.
The second motivation for this representation is that it enables the use of the multiple-scattering translation that we will apply to accelerate the solution process. Thirdly, cylindrical harmonics are easily rotated and thus only one transformation needs to be calculated for inclusions that are identical up to rotation. Nonetheless, it is difficult to ascertain \emph{a priori} what the optimal number of cylindrical harmonics is for a given inclusion in a multiple-scattering problem, as this number depends not only on the type and frequency of the incident wave but also on the shape of the inclusion and the distance between it and its closest neighboring inclusion. In the past few years some convergence bounds have been developed \cite{ar:Ganesh2012}, but in our examples these proved to be highly shape-dependent and not as accurate in the near field, and therefore we relied on a computational approach to determine the optimal number.

One drawback of this transformation is its inability to handle touching or intersecting scattering disks, which are fictitious circles strictly enclosing the inclusions, even if the inclusions themselves are adequately separated. The worst manifestation of this issue would occur with thin and long inclusions whose scattering disks cover a disproportionately large area. However, one can partially overcome this restriction by grouping multiple inclusions in close proximity into one disk and rotating them in unison.

We utilize a layer potentials formulation~\cite{ar:rokhlin83}, wherein a single-layer potential density $\sigma$ and a double-layer potential density $\mu$ are assumed to exist on $\partial \Omega$. For notational simplicity, in this section we assume that the inclusion surface $\Omega$ is centered at the origin. Note that although we focus only on smooth shapes, if $\partial \Omega$ is not smooth, the method is still applicable with an appropriate discretization approach~\cite{ar:bremer2010}.
These densities have unknown complex amplitudes and give rise to the potential representation
\begin{align}\label{eq:usc_operator}
\usc = \begin{cases}
\mathcal{S}^{\kin}\sigma + 
\mathcal{D}^{\kin} \mu & \mbox{in } \Omega, \\
\mathcal{S}^{\kout} \sigma + 
\mathcal{D}^{\kout} \mu & \mbox{otherwise}
\end{cases}
\end{align}
for the $\hat{\Vz}$ component of the scattered electric field, where the single- and double-layer potential operators for wavenumber $k$ are defined by
\begin{align}
\begin{array}{lll}
\mathcal{S}^{k} \sigma(\Vr) &:=& \displaystyle{\int_{\partial \Omega} G^{k} (\Vr,\Vr') \sigma(\Vr') \dif \Vr'},
\\
\mathcal{D}^{k} \mu(\Vr) &:=& \displaystyle{\int_{\partial \Omega} \dpd{G^{k}}{n_{\Vr'}}  (\Vr,\Vr') \mu(\Vr') \dif \Vr'}
\end{array}
\end{align}
and $G^{k} (\Vr,\Vr') = \frac{i}{4}H_0^{(1)} (k|\Vr-\Vr'|)$ is the two-dimensional Green's function for the Helmholtz equation in a homogeneous material. For a given incident field $\uinc$, the constant-permeability TM$_z$ boundary conditions are applied to the potential formulation. After accounting for the potential density jump across the boundary~\cite{ar:kress94} we have the system
\begin{align}
\begin{array}{lll}
\mathcal{S}^{\kout} \sigma - \mathcal{S}^{\kin} \sigma + \mathcal{D}^{\kout} \mu - \mathcal{D}^{\kin} \mu + \mu & =& -\uinc,
\\
\dpd{}{n_\Vr}
\left(
\mathcal{S}^{\kout} \sigma - 
\mathcal{S}^{\kin} \sigma \right)
+ 
\dpd{}{n_\Vr}
\left(
\mathcal{D}^{\kout} \mu - 
\mathcal{D}^{\kin} \mu \right) - \sigma &=& -\dpd{\uinc}{n_\Vr}
\end{array}
\end{align}
of integral equations which holds for all points $\Vr \in \partial \Omega$.
This system cannot be solved by directly evaluating the operators on the boundary on account of the singularity in $G^k$ and the hypersingularity in its second-order derivative. Hence we split each integrand into two terms~\cite{ar:kress94}, integrating the first term with the Kussmaul-Martensen quadrature rule and the other with trapezoidal or Gauss-Legendre quadrature. Many other choices for the quadrature rule exist and can be used interchangeably, such as the more sophisticated QBX~\cite{ar:qbx13}.
Denote the values of the potential densities $\sigma$, $\mu$ on $2N$ discretization nodes by $\Vsigma$, $\Vmu$ respectively. We obtain the system of equations 
\begin{align}
\label{eq:density_system}
\VA
\Bigg(
\begin{matrix}
\Vsigma \\ \Vmu
\end{matrix}
\Bigg)
= -
\Bigg(
\begin{matrix}
\uinc \\ \pd{\uinc}{n}
\end{matrix}
\Bigg)
,
\end{align}
in which $\VA$ is a $4N \times 4N$ matrix which includes all potential operators.

In order to expand the potentials in terms of cylindrical harmonics, the system in Eq.~(\ref{eq:density_system}) is solved for $2P+1$ incoming waves sampled on the discretization points of the shape, or $\uinc = J_p(\kout|\Vz|)e^{i p \angle \Vz}$ for $p=-P, \dots, P$. This yields the single- and double-layer potential density vectors $\Vsigma_p$, $\Vmu_p$ for the $p$-th incident wave. For this solution method to maintain reasonable time complexity, this system should be factorized (e.g.~LU) for successive direct solutions, thus requiring $O(N^3 + (2P+1)N^2)$ computations in total. 

Let $\Vr$ be a point that lies strictly outside the inclusion such that $| \Vr | > | \Vr'|$ for any $\Vr'$ on the boundary. We apply Graf's addition theorem for Hankel functions to the integral operator formula for the scattered field given by Eq.~(\ref{eq:usc_operator}) and obtain the cylindrical harmonics expansion
\begin{align}
\label{eq:expansion}
&\usc(\Vr) = \sum_{l=-P}^{P} s_{l,p} H_l^{(1)}(\kout |\Vr|)e^{i l \angle \Vr},  \notag
\\
&s_{l,p} := \frac{i}{4} \int_{\partial \Omega} J_l(\kout|\Vr'|) e^{-il\angle \Vr'}\sigma_p(\Vr') +
\mathbf{\hat{n}}_{\Vr'} \cdot \nabla \pb{J_l(\kout|\Vr'|)e^{-il\angle \Vr'}}\mu_p(\Vr')
\dif \Vr'
\end{align}
of the potential operators.
Notably, this expansion only holds strictly outside the inclusion, and thus we assume a fictitious scattering disk $D$ which strictly encloses the inclusion. Inside this disk, the direct integral equation representation is assumed, while outside of it the expansion in Eq.~(\ref{eq:expansion}) holds. In this work the diameter of the scattering disks is chosen to be $10\%$ larger than the inclusion diameter. While the diameter of the scattering disk can be reduced if necessary, this typically leads to a dramatic increase in $P$. Approximating the integral above with the same boundary discretization yields a formula of the form $s_{l,p} = (\mathbf{A} {\Vsigma}_p + \mathbf{B} {\Vmu}_p)_l$,
which in turn yields the entire scattering matrix 
$\mathbf{X}^{(m)} = \VA \VSigma + \VB \VM$
for the $m$-th inclusion,
where the $p$-th column of $\VSigma$ is $\Vsigma_p$ and similarly for $\VM$ and $\Vmu_p$.

As mentioned earlier, the process above only needs to be carried out once per inclusion, up to rotation. The representation of an inclusion rotated by an angle $\varphi_m$ is readily available by multiplying the ($l,p$)-th element of $\VX^{(m)}$ by a factor of $e^{-i\varphi_m(l-p)}$, in other words, by replacing the scattering matrix with $\VPhi \VX^{(m)} \VPhi^*$ for the diagonal matrix $\VPhi_{p,p} = e^{-ip\varphi_m}$.

Now let $\Omega_m$ be centered at $\Vo^{(m)}$ with a local coordinate system $\Vr^{(m)} = \Vr -\Vo^{(m)}$.
In order to use the scattering matrix to solve scattering of an incident field $\uinc$ from the single $m$-th inclusion, we first expand $\uinc$ as
\begin{align}
\uinc = \sum_{p=-P}^{P}
\alpha_p^{(m)} J_p(\kout |\Vr^{(m)}|) e^{i p \angle \Vr^{(m)}}.
\end{align}
Due to the Jacobi-Anger expansion in the particular case of plane-wave incidence $e^{i\mathbf{k} \cdot \Vr}$ for some
$\mathbf{k} = (k \cos \theta_i, k \sin \theta_i)$, we have $\alpha_p = e^{i p \left(\pi/2 - \theta_i \right)}$ in the local coordinates up to multiplication by a phase constant.
The electric field scattered by the inclusion is given by the outgoing expansion
\begin{align}
\label{eq:u_sc_beta}
\usc = 
\sum_{p=-P}^{P}
\beta_p^{(m)} H_p^{(1)} (\kout |\Vr^{(m)}|) e^{i p \angle \Vr^{(m)}},
\end{align}
that is, a linear combination of the scattering matrix columns, where in this case, $\beta_p^{(m)} = (\VX^{(m)} \Valpha^{(m)})_p$.
Note that circular inclusions can be analytically represented using a diagonal scattering matrix by utilizing orthogonality of the basis functions on a circle. For such an inclusion with radius $R$, the scattering matrix components are readily given by
\begin{align}
\beta_p &= -\alpha_p
\frac{J_p(\kout R)J_p'(\kin R) - J_p'(\kout R)J_p(\kin R)}
{H_p^{(1)}(\kout R)J_p'(\kin R) - H_p^{(1)\prime}(\kout R)J_p(\kin R)},
\end{align}
where $Z_p'(kR) = k \left( Z_{p-1}(kR) - (p/{kR}) Z_{p}(kR) \right)$ for $Z_p = J_p,H_p^{(1)}$.

\begin{figure}[htb]
\centering
\includegraphics{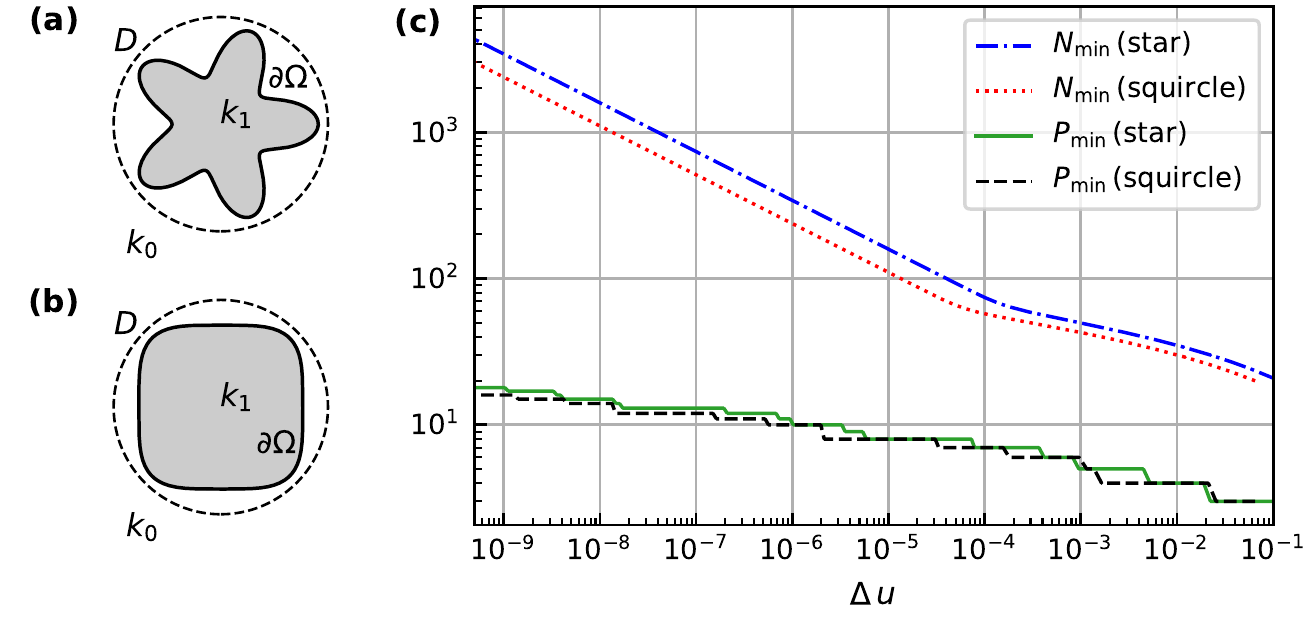}
\caption{Minimum discretization nodes and cylindrical harmonics for two inclusion shapes.
(a) Rounded star with the smooth boundary parametrization
$\Vr(\theta) = \pb{R + a \cos(5\theta)}\left(\cos \theta, \sin \theta\right)$ for $R=0.3\lambda_0$, $a=0.1\lambda_0$, and $k_1=1.5k_0$, and its scattering disk $D$. 
(b) Squircle with $R=0.35 \lambda_0$ and $k_1=1.5k_0$, and its scattering disk $D$. 
(c) Minimum values of the discretization nodes $N$ and number of cylindrical harmonics $P$ for given discretization and cylindrical transformation error, respectively, for these two inclusions.}
\label{fig:P_error_example}
\end{figure}

Two error mechanisms affect the accuracy of the solution beyond the adjustable FMM truncation and quadrature error discussed in Section~\ref{ssec:FMM}.
First we have the discretization error due to the finite number of nodes $2N$ on the inclusion boundary, and the second stemming from the transformation to a cylindrical harmonics formulation.
We denote by $\Delta u$ the normalized RMS errors for these error mechanisms.
The discretization error is computed as follows: a fictitious line source is assumed at some point inside the inclusion along with an incident plane wave outside of it. The potential densities $\Vsigma$, $\Vmu$ on the boundaries $\partial \Omega$ attained from solving the potential density system of Eq.~(\ref{eq:density_system}) induce fields outside the inclusion that are equivalent to those of the line source, up to the error that is measured on the scattering disk $D$.
The cylindrical harmonics transformation error is measured by comparing the field induced by the potential densities to that of the cylindrical harmonics on points distanced $2D$ from the inclusion center.
Fig.~\ref{fig:P_error_example} shows an example of the relation between $N$ and $P$ and their respective errors for two inclusion shapes.
Note that not only is $N$ substantially larger than $P$ for all values of $\Delta u$, but the ratio between them continues to grow as the desired errors diminish.

\subsection{Multiple-scattering formulation}
\label{sec:method_multiple}
Here we apply the principles used in the preceding section to a multiple-scattering setting. Previously, the relation between incoming and outgoing coefficients was given by the scattering matrix, however, the incident field of a single inclusion in a multiple-scattering scenario is a combination of the incident field and the fields reflected off all other inclusions. A translation matrix is used to transform the reflected field from the local coordinates of one inclusion to the local coordinates of another~\cite{ar:gumerov2007}.

Let $\Vr^{(m)}$ and $\Vr^{(m')}$ denote a point in the local coordinates of the $m$-th and $m'$-th inclusions, respectively, and let $\Vr^{(m',m)}$ be the coordinates of the $m'$-th inclusion with respect to the center of the $m$-th inclusion. Using Graf's addition formula and truncating the higher-order elements, we obtain the relation between the two local expansions
\begin{align}
\sum_{p=-P}^{P} \beta_p^{(m)} H_p^{(1)} (\kout |\Vr^{(m)}|)e^{i p \angle \Vr^{(m)}} = 
\sum_{\mu=-P}^{P}
J_\mu (\kout |\Vr^{(m')}|) e^{i \mu \angle \Vr^{(m')}}
\sum_{p=-P}^{P} \beta_p^{(m)}
\left( \VT^{(m',m)} \right)_{\mu,p}
\ ,
\end{align}
where $\VT^{(m',m)}$ with the elements
\begin{align}\label{eq:translation}
\left(\VT^{(m',m)}\right)_{\mu,p} = e^{i (p-\mu) \angle \Vr^{(m',m)}}
H_{p-\mu}^{(1)} (\kout \Vr^{(m',m)})
\end{align}
is the translation matrix which translates the outgoing coefficients of one inclusion to the incoming coefficients of another.
Summing over the contributions of all the inclusions, we obtain the complete incoming coefficients of the $m'$-th particle
\begin{align}\label{eq:multiple_scattering_incomplete}
{\tilde{\Valpha}}^{(m')} = \Valpha^{(m')} + \sum_{m \neq m'} \VT^{(m',m)} \Vbeta^{(m)} .
\end{align}
Finally, we note that $\Vbeta^{(m')} = \VX^{(m')} \tilde{\Valpha}^{(m')}$ holds for the complete incoming coefficients, and substitute this relation into Eq.~(\ref{eq:multiple_scattering_incomplete}) to obtain
\begin{align}
\left(\VX^{(m')}\right)^{-1} \Vbeta^{(m')} -
\sum_{m \neq m'} \VT^{(m',m)} \Vbeta^{(m)} 
= \Valpha^{(m')}
,
\end{align}
thus yielding a system of $(2P+1)M$ equations, where $M$ is the number of inclusions. A preconditioned scattering system is obtained when multiplying both sides by the block scattering matrix, which we denote in concatenated form by
\begin{align} \label{eq:IXT}
\left(\VI - \VX \VT\right) \Vbeta = \VX \Valpha .
\end{align}

Once the multiple-scattering system in Eq.~(\ref{eq:IXT}) is solved, the scattered field at any point outside the scattering disks is readily calculated by summing Eq.~(\ref{eq:u_sc_beta}) over all inclusions. Strictly inside the inclusions, the field is given by the discretized integral Eq.~(\ref{eq:usc_operator}), where the densities are
\begin{align}
\Vsigma^{(m)} = \VSigma \left(\VX^{(m)}\right)^{-1} \Vbeta^{(m)}
,\quad
\Vmu^{(m)} = \VM \left(\VX^{(m)}\right)^{-1}  \Vbeta^{(m)}.
\end{align}
These are weighted sums of those $\Vsigma_p$, $\Vmu_p$ obtained from solving Eq.~(\ref{eq:density_system}) for the different incoming waves, as the expansion in Eq.~(\ref{eq:u_sc_beta}) of the inclusion is not valid inside the scattering disk. 
Between the $m$-th inclusion and its disk, the scattered field $\usc$ is given by summing Eq.~(\ref{eq:u_sc_beta}) over all $m'\neq m$ and then adding the direct integral operator for $m$.

\subsection{FMM acceleration of the translation} \label{ssec:FMM}
As the computational cost of directly solving Eq.~(\ref{eq:IXT}) becomes prohibitively high for a large number of inclusions, this system should be solved iteratively.
While applying the block-diagonal scattering matrix $\VX$ in each iteration requires only $O(M)$ operations, the translation matrix is almost fully populated and thus requires $O(M^2)$ operations. Therefore, we choose to apply the block translation matrix $\VT$ using FMM~\cite{ar:coifman1993}, yielding a lower complexity that will be analyzed in the next section. 
In this section, we shall succinctly describe the FMM process for this problem. Assume a collection of many inclusions, divided into $G$ non-empty $a \times a$ boxes. The FMM process converts the translation matrix to a sequence of operators. These operators aggregate the translation matrices of multiple inclusions in one box, translate them to a different box and disaggregate them to the inclusions in said box. Note that this process assumes the boxes have some minimal distance between them. For boxes which are closer than this minimal distance, or are the same box, the appropriate blocks of the translation matrix $\VT$ are directly applied via a sparse near-interaction matrix.

Let the $m$-, $m'$-th inclusions which are centered at $\Vo^{(m)}$, $\mathbf{o}^{(m')}$ be placed in boxes centered at $\Vc$, $\Vc'$ respectively. Provided $\Vc$, $\Vc'$ are distanced by at least $\sqrt{2}a$, Graf's addition and Bessel's integral theorems are applied to the translation matrix in Eq.~(\ref{eq:translation}), which results in
\begin{align}
\left(\VT^{(m',m)}\right)_{\mu,p} =
\frac{1}{2\pi}
\int_0^{2\pi}
e^{i\Vk \cdot (\Vo^{(m')} - \Vc')}
\mathcal{F}_\infty(\theta,\Vc'-\Vc)
e^{-i \Vk \cdot (\Vo^{(m)} - \Vc)}
e^{i\left(\mu-p\right)\left(\pi/2-\theta\right)}
\dif \theta
,
\end{align}
where $\Vk = (k\cos\theta,k\sin\theta)$, and the truncated FMM translation function which transmits plane waves from $\Vc$ to $\Vc'$ is defined as
\begin{align}
\mathcal{F}_{P_{\mathrm{FMM}}} (\theta,\Vx) :=
\sum_{\xi=-{P_{\mathrm{FMM}}}}^{P_{\mathrm{FMM}}}
H_\xi^{(1)}(k| \Vx|) e^{i\xi \left(\angle \Vx + \pi/2 - \theta\right)}.
\end{align}

Although this translation function must be truncated for practical computations, the series does not converge for small values of $P_{\mathrm{FMM}}$ and oscillates for large values, making the optimal choice of an extensively-studied, non-trivial problem. Several analytical and empirical formulas have been proposed for this truncation, of which the excess bandwidth formula~\cite{bk:chew2001fast} is used here. Assuming this series truncation, the integral expansion of the Bessel function has finite bandwidth such that a $Q \propto P_{\mathrm{FMM}}$-point quadrature of $[0,2\pi]$ is sufficient. Hence if we define $\Vk_q := (k\cos\theta_q,k\sin\theta_q)$, the translation matrix is approximated as
\begin{align}
\left({\VT}^{(m',m)}\right)_{\mu,n} \approx
\frac{1}{Q}
\sum_{q=1}^{Q}
\underbrace{e^{i \Vk_q \cdot (\Vo^{(m')} - \Vc')}
e^{i \mu \left(\pi/2-\theta_q\right)}}_{\textrm{disaggregation}}
\mathcal{F}_{P_{\mathrm{FMM}}} (\theta_q,\Vc'-\Vc) 
\underbrace{e^{-i \Vk_q \cdot (\Vo^{(m)} - \Vc )}
e^{-i n \left(\pi/2-\theta_q\right)}}_{\textrm{aggregation}} .
\end{align}

We now construct the FMM matrices used for matrix-vector product acceleration. Denote by $M_g$ the number of inclusions in the $g$-th box, centered at $\Vc_g$. We construct a $1 \times M_g$ block aggregation matrix, containing a block for every inclusion, with the $m$-th block given by
\begin{align}\label{eq:aggregation}
\left({\VA}^{(m)}\right)_{q,n} :=
e^{-i \Vk_q \cdot (\Vo^{(m)} - \Vc_g) - in \left(\pi /2 - \theta_q\right)}, \quad  q = 1, \dots, Q, \quad n = -P, \dots, P
\end{align}
Since FMM is applied to every box with respect to every other box, we construct the disaggregation matrix by applying the conjugate transpose to the aggregation matrix. 

Finally, for each pair $(g',g)$ of sufficiently distant boxes, a diagonal FMM translation matrix $\VF^{(g',g)}$ is constructed by
\begin{align}
\left({\VF}^{(g',g)}\right)_{q,q} := \frac{1}{Q} \mathcal{F}_{P_{\mathrm{FMM}}} (\theta_q,\Vc_{g'} - \Vc_g), \quad q=1, \dots, Q .
\end{align}

\subsection{FMM complexity}\label{ssec:complex}
Complexity analyses for the application of the FMM to various problems are well established, generally leading to a single-level result of $O( N_{\mathrm{dof}}^{1.5})$ and multi-level complexity $O( N_{\mathrm{dof}} \log N_{\mathrm{dof}})$ for $N_\mathrm{dof}$ degrees of freedom. However, the relationship between the optimal number of boxes and the wavenumber is different in the multiple-scattering approach, and therefore we find it instructive to briefly analyze the complexity of our FMM application.

Since each aggregation matrix is of dimension $Q \times M_g(2P+1)$, performing the aggregation of all $G$ boxes has time complexity $O \left(MQ\left(2P+1\right)\right)$, and thus so does the total disaggregation. The time complexity of performing all box-to-box FMM translations is $O(QG^2)$, while the number of nonzero elements in the near-interaction matrix is
\begin{align}
(2P+1)^2 \Big[ \sum_g M_g (M_g - 1) + \sum_g M_g \sum_{(g',g) \textrm{ near}} \!\!\!\! M_{g'} \Big]
.
\end{align}

Therefore, applying the near-interaction matrix is expected to require $(2P+1)^2 \sum_g [M_g^2 + M_g]$ operations. Including the computational cost of applying the scattering and identity matrices, applying the operator $(\VI-\VX \VT)$ using FMM has time complexity
\begin{align}
O \Big(
MQ(2P+1)
+
QG^2
+
(2P+1)^2 \sum_g [ M_g^2 + M_g ]
+
M(2P+1)^2
\Big).
\end{align}
Since the quadrature $Q$ is proportional to the diameter of each box, and in two dimensions the area of a box is inversely proportional to the number of boxes, we have $Q \propto G^{-0.5}$. If we assume an approximately constant distribution of inclusions in boxes such that $M_g \approx M/G$, the FMM time complexity expression is simplified to
\begin{align}
O \left(
G^{1.5}
+
(2P+1)^2 M^2 G^{-1}
\right).
\end{align}

We note that while the usual FMM choice $G \propto \sqrt{M}$ yields a complexity of $O(M^{1.5})$, selecting $G = bM^{0.8}$ for a constant $b$ reduces the complexity to $O(M^{1.2})$ per FMM solution with regard to the number of inclusions. In practice, even a choice of $G \propto M$ may be optimal due to the quadratic dependence of the second complexity term on the wavelength. An analogous analysis of a Multi-Level Fast Multipole Algorithm approach will lead to asymptotic complexity of $O(M)$ \cite{ar:lai14}, although this is only beneficial in practice for very large values of $M$.

\section{Optimization for multiple-scattering features} \label{sec:optimization}
We give a description of a general optimization problem that is applicable to various metamaterials, where our aim is to provide a template for applying our framework to different devices. 
Given an objective function as in Eq.~(\ref{eq:fobj}), we develop its gradient, and show how it can be computed in order to find optimal parameters for the overall structure. 
Our $I$ points of interest $\Vr_i$ are assumed to lie outside all scattering disks, as points inside them complicate and slow down the optimization procedure.
Note that whether we are minimizing or maximizing the objective function is immaterial, as maximization problems can be solved by minimizing the negated objective function and again negating the achieved minimum value. Simultaneously minimizing intensity at several points while maximizing it at others is achieved by appropriately weighting the objective function.
For convenience, we rewrite the field values in the objective function in terms of $\Vbeta$ and obtain the column vector $\Vu  = \VH^T \Vbeta + \Vu^{\mathrm{inc}}$ and the simplified form $f_{\mathrm{obj}} = \| \Vu \| ^2$, where $\VH$ relates the coefficient solution to the objective function. 

Let $\Vw$ denote a vector of $J$ inclusion parameters, where we assume each parameter affects the shape of an inclusion, but not the location of its center, and therefore $\VH$ remains constant.
In order to calculate the gradient $\nabla f_{\mathrm{obj}}$ with respect to $\Vw$, we shall use the adjoint-state method~\cite{ar:chavent1974identification, ar:plessix2006adjoint}, as its complexity is less dependent on the number of design variables than a direct approach. Our optimization problem is given by
\begin{align}
\begin{array}{lll}
\displaystyle\min_{\Vw} & & f_\mathrm{obj}(\Vbeta) = \|\VH^T \Vbeta + \Vu^\mathrm{inc}\|^2
\\
\text{subject to} & & \Vc(\Vbeta,\Vw) = \pb{\VI - \VX(\Vw) \VT} \Vbeta - \VX(\Vw) \Valpha = 0
\end{array}
\end{align}
To apply the adjoint-state method, we utilize a complex vector $\Vlambda$ to define the Lagrangian
\begin{align}
\Lambda = f_\mathrm{obj} + \Vlambda^T \Vc + \overbar{\Vlambda^T \Vc},
\end{align}
equate the complete derivatives of $f_\mathrm{obj}$ and $\Lambda$ with respect to $w \in \Vw$, and have after some algebraic manipulation that
\begin{align}\label{eq:adjoint}
\frac{\mathrm{d} f_\mathrm{obj}}{\mathrm{d} w}
=
2\Re \left\lbrace \pb{\dpd{f_\mathrm{obj}}{\Vbeta} + \Vlambda^T \left(\VI - \VX \VT\right)}\dpd{\Vbeta}{w} \right\rbrace
 - 2\Re \left\lbrace \Vlambda^T \dpd{\VX}{w} \VX^{-1} \Vbeta \right\rbrace
.
\end{align}
The crux of the adjoint-state method resides in setting the first summand to zero by properly solving for $\Vlambda$. This will allow us to calculate the derivative without explicitly computing $\partial \Vbeta / \partial \Vw$ which would add significant complexity. Substituting $f_\mathrm{obj}$ yields the adjoint system
\begin{align}\label{eq:adjoint_system}
\left(\VI - \VT^T \VX^T\right) \Vlambda = - \VH \overbar{\Vu},
\end{align}
which we solve using a modified FMM procedure with the same complexity.
Once the system is solved, each element of the gradient can be calculated in $O((2P+1)^2)$ time, yielding $O(M(2P+1)^2) + O(\text{FMM})$ complexity in total if each inclusion is affected by a single parameter.
A description of the complete process of automatically designing a device via our approach is summarized in Algorithm~\ref{algo:1}. The specifics depend on the optimization method used, where additional evaluations of $f_\mathrm{obj}$ might be necessary for the optimization line search.
We note that $\VX^{-1}$ was computed in a previous step and its use here is not problematic, and in any event $\VX^{-1}\Vbeta$ can be replaced with $\VT\Vbeta + \Valpha$.

\begin{center}
\begin{algorithm}[H]
$\Vw \gets (w_1,\dots,w_J)$ \tcp{initial value for optimization}
\tcp{Precomputation phase}
\For{all distinct non-circular inclusions}{
Construct and solve potential density equation~(\ref{eq:density_system}) for $-P,\dots,P$
}
Prepare FMM matrices \tcp{using the development in Section~\ref{ssec:FMM}}
\While{optimization has not converged}{	
	$\Vbeta \gets$ solution of multiple-scattering equation~(\ref{eq:IXT}) with FMM
	
	Calculate $f_\mathrm{obj}$ using $\Vbeta$

	\tcp{Construct gradient:}
	
	Solve adjoint system of Eq.~(\ref{eq:adjoint_system}) for $\Vlambda$ using adjoint FMM
	
	\For{$w_j \in \Vw$}{
		Compute $j$-th component of $\nabla f_\mathrm{obj}$ using Eq.~(\ref{eq:adjoint})
	}
	$\Vw \gets$ next optimization point
}
\caption{Automated design of dielectric metamaterials}
\label{algo:1}
\end{algorithm}
\end{center}

In this work, we optimize inclusion parameters for which $\nabla_{\Vw} \VX$ is analytic, such as the rotation angle of an arbitrary inclusion and the radius of a circular inclusion, which significantly simplifies the computation of the gradient. Attempting to optimize parameters that do change the structure of $\VX$ is more involved, and may require numerical differentiation.

\section{Numerical results}
\label{sec:results}
In this section, we demonstrate our approach using three examples. First, we study the run time of the multiple-scattering approach for increasingly numerous inclusions. Additionally, we apply the optimization process in its entirety to two practical examples, resulting in improved designs.
In what follows, all values of $2N$, the number of discretization nodes, and $P$, the cylindrical harmonics parameter, are chosen to be the minimal values for which an electric field error of $10^{-6}$ holds, as explained in Section~\ref{sec:math}. All linear systems solved via FMM use GMRES~\cite{ar:saad1986gmres} with tolerance $10^{-6}$ as the underlying iterative method.
All simulations were written in the Julia programming language~\cite{ar:julia2017review}, and run on a 3.4GHz Intel Core i7-6700 CPU with 32GB of memory.

\subsection{Complexity of multiple-scattering approach}\label{ssec:numerical1}

\begin{figure}[htb]
\centering
\includegraphics{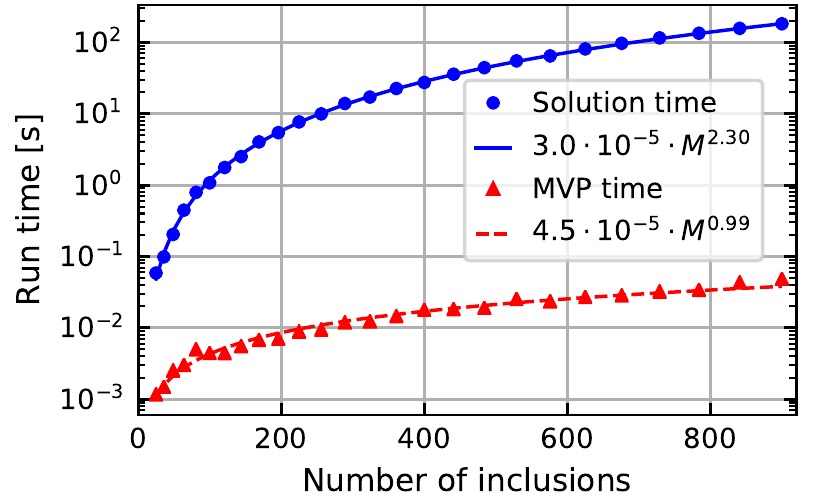}
\caption{Run time of the multiple-scattering system solution, as well as of a single matrix-vector product, for different numbers of inclusions. Here an incident plane wave is scattered by a $\sqrt{M} \times \sqrt{M}$ grid of identical rounded stars, randomly rotated. The inclusion parameters are $R = 0.3\lambda_0$, $a=0.1 \lambda_0$ and $k_1 = 1.5 k_0$, and are distanced $0.9\lambda_0$.}
\label{fig:sim_complexity}
\end{figure}

We examine the run time of the multiple-scattering algorithm for a square $\sqrt{M} \times \sqrt{M}$ grid of inclusions, and compare it to the theoretical complexity analysis in Section~\ref{ssec:complex}.  
Fig.~\ref{fig:sim_complexity} depicts the run time of solving the multiple-scattering Eq.~(\ref{eq:IXT}) using FMM for several values of $M$. 
The minimal values of $N$ and $P$ for $\Delta u = 10^{-6}$ and this inclusion are $N=342$ and $P=10$. The precomputation of the prototype inclusion for these values was performed once for all simulations and required $0.9\,\mathrm{s}$ that were not included in the plot.
A single matrix-vector product scales almost linearly with the number of inclusions, in accordance with the complexity analysis. The total solution convergence time has complexity $O(M^{2.3})$, i.e., the number of iterations depends on the number of inclusions, which is not uncommon when solving electromagnetic equation systems with Krylov subspace methods. Nonetheless, the total solution time is several orders of magnitude below that achievable by a naive method.

\subsection{Rotation-angle optimization for arbitrary inclusions}
For our first optimization example, we apply our framework to the optimization of inclusion rotation. That is, given an incident wave with wavelength $\lambda_0$ scattered by a collection of $M$ inclusions, we wish to find the optimal rotation angles $\Vvarphi$ of the inclusions such that the field propagation in some desired direction is maximized.

\begin{figure}[htb]
\centering
\includegraphics{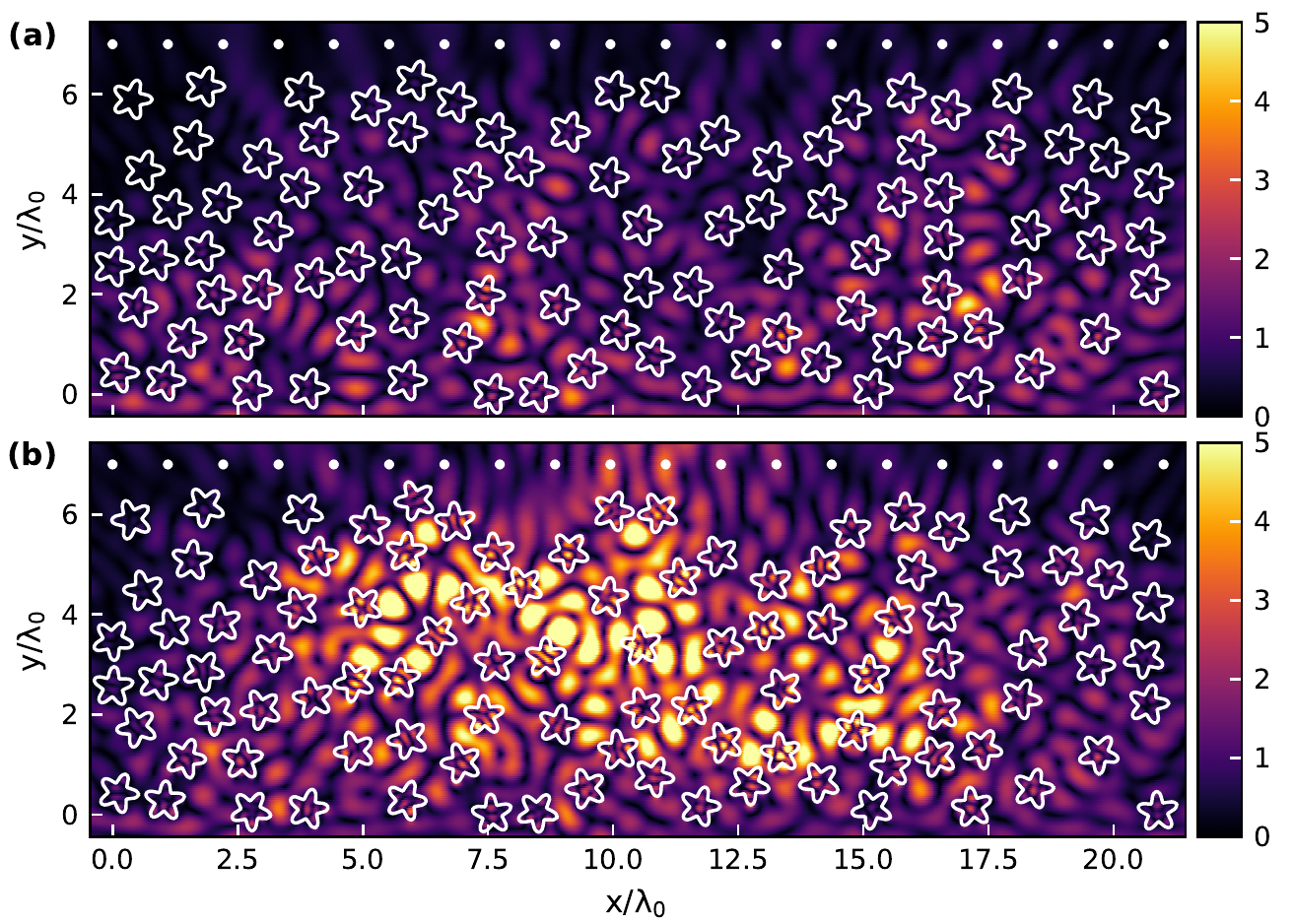}
\caption{Optimization of rotation angles. 
(a) Initial electric field amplitude after scattering by $M = 100$ randomly positioned identical rounded stars with zero rotation, which prevent the $\hat{\Vy}$-traveling plane wave from propagating in its original direction.
(b) Electric field amplitude for the same inclusions, with rotation angles optimized to maximize field at $20$ points along the top boundary. Markers indicate points where the field is maximized.}
\label{fig:opt_angle_pt1}
\end{figure}

\begin{figure}[htb]
\centering
\includegraphics{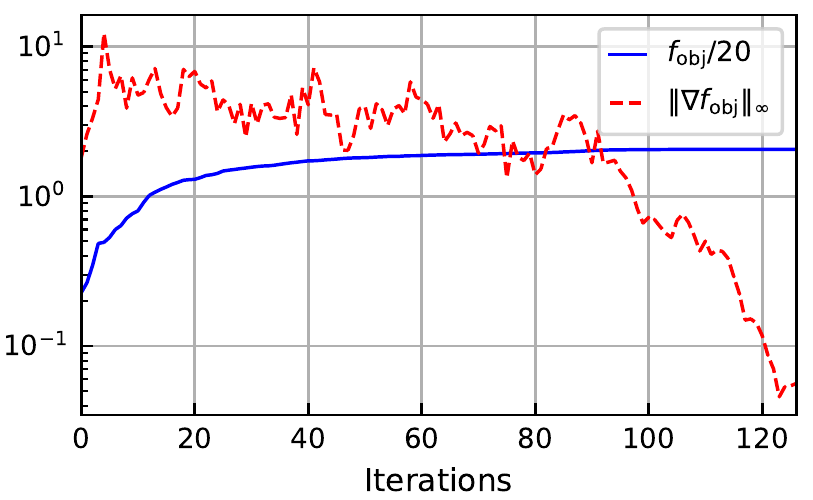}
\caption{Convergence behavior of the objective function $f_{\mathrm{obj}}$ and its gradient $\nabla f_{\mathrm{obj}}$ for Fig.~\ref{fig:opt_angle_pt1}.}
\label{fig:opt_angle_pt2}
\end{figure}

The derivatives of the scattering matrices with respect to the rotation angles are given by
\begin{align}
\left( \dpd{{\VX ^{(m)}}}{\varphi_j}\right)_{u,v} = - i \delta_{m,j} (u-v) \left(\VX^{(m)}\right)_{u,v} = \delta_{m,j} \left(\VD \VX^{(m)} - \VX^{(m)} \VD\right)_{u,v}
,
\end{align}
where $(\VD)_{u,v} = -\delta_{u,v} i u$. 
Since the rotation angles are unconstrained, our choice of optimization method is the Broyden-Fletcher-Goldfarb-Shanno (BFGS)~\cite{bk:optimization} algorithm, which is a quasi-Newton method that locally approximates the objective function as a quadratic. In each iteration, once the descent direction is decided via the gradient, a line search is necessary to determine the step size to the minimum in that direction. The backtracking line search based on the Armijo-Goldstein condition~\cite{ar:armijo}, which minimizes gradient evaluations, is used here.
In Fig.~\ref{fig:opt_angle_pt1}, we simulate the case of a $\hat{\Vy}$-traveling plane wave incident upon a collection of $M=100$ inclusions, randomly positioned in a $21\lambda_0 \times 7\lambda_0$ rectangle such that the scattering disks do not intersect.
Inclusions are rounded stars with the same size as in Fig.~\ref{fig:P_error_example}, have wavenumber $k_1 = 3k_0$ and use the minimal parameters $N=934$, $P=12$. 
The objective function is set as in Eq.~(\ref{eq:fobj}) for $I=20$ points of interest $\Vr_i$ located equidistantly along the top boundary of the rectangle, which are indicated with white dots.
The field amplitude at the points of interest $\Vr_i$ is substantially larger after the optimization process, whose convergence is shown in detail in Fig.~\ref{fig:opt_angle_pt2}.
Specifically, the BFGS method converges to an average field magnitude of $1.43$ at $\Vr_i$, up from the initial value of $0.48$ (in the RMS sense), a $200\%$ increase. The process required $127$ iterations and $664$ seconds for the convergence criterion $\Delta f_{\mathrm{obj}} < 10^{-6}$.

\subsection{Radius optimization for circular inclusions}\label{ssec:numerical3}
We now consider optimization of the radii of circular inclusions, where in contrast to the previous example, both the scattering matrices and their derivatives with respect to the inclusion radius are diagonal and have analytical form.
This example is motivated by the photonic crystal implementation of the Luneburg lens.
The two-dimensional Luneburg lens~\cite{bk:optics} is a symmetric circular lens designed such that incoming plane waves are focused to a single point on its rim, and no waves are reflected. This property is achieved by a continuously varying refractive index given by the analytic solution $n(r) = \sqrt{2-(r/R_{\textrm{lens}})^2}$, where $r$  is the distance from the center of the lens, which has radius $R_{\textrm{lens}}$. One way of fabricating a Luneburg lens is via long dielectric rods on a glass substrate, which, if long enough, can be assumed to be infinite. Thus the electromagnetic propagation through the device can be treated as a two-dimensional problem. In this setting, the lens is divided into unit cells on a square grid, each with side length $a$. Each unit cell $m$ contains a circular inclusion with the same relative permittivity $\varepsilon_r$ but differing radius $R_m$, such that the effective refractive index in the cell can be approximated analytically if $a/\lambda_0$ is sufficiently small~\cite{ar:luneburg2010}, and thus the radii are set such that the average permittivity approximates the Luneburg solution.
 
This implementation of the Luneburg lens begs the question whether the electromagnetic focusing could be improved by sacrificing the rotational symmetry of the device, however, note that the restriction to a square grid has already limited this symmetry. 
To answer this question, we propose optimizing over the radii of the inclusions to maximize the field amplitude at the focal point. 
Note that since the inclusions are circular, the computation of the gradient is cheaper than in the previous example, as is applying the diagonal scattering matrix in each FMM iteration. Care must be taken to assure that the computed radii are neither below some non-negative lower practical limit $R_{\mathrm{min}}$ nor above the limit $R_{\mathrm{max}}$ at which they are too close for the multiple-scattering approximation in this work, i.e.~$0.45a$. 
Thus unconstrained optimization methods such as BFGS are no longer an option. Fortunately, these so-called box constraints are simple enough to be tackled by the addition of a penalty term which sharpens the constraint from one BFGS run to the next.

\begin{figure}[htb]
\centering
\includegraphics{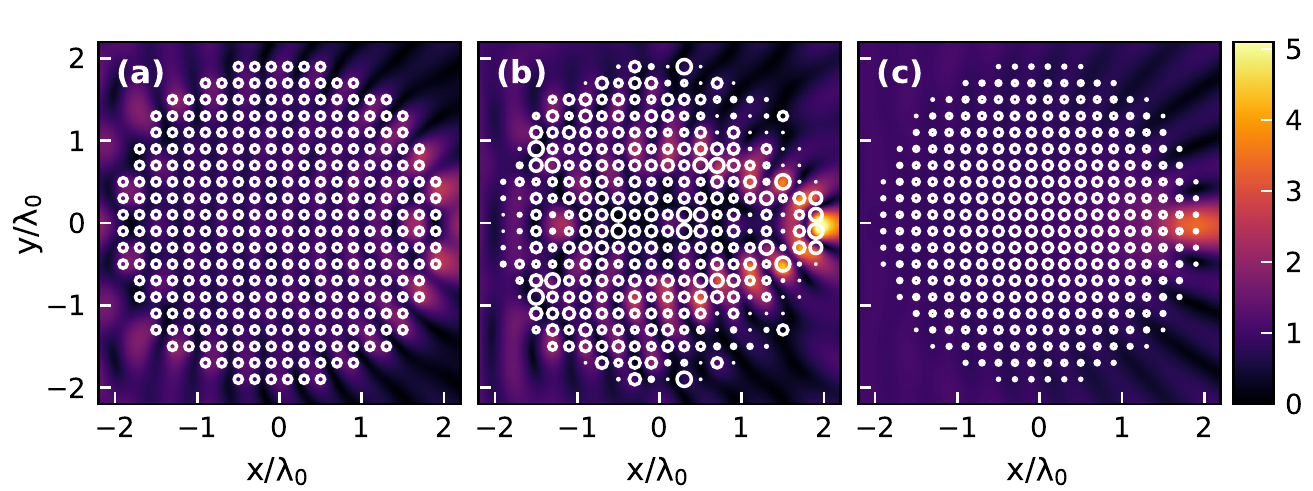}
\caption{Radius optimization of 316 circular inclusions with $\varepsilon_r=4.5$ for focusing an $\hat{\Vx}$-traveling plane wave to a single focal point on the lens rim. Electric field amplitude for 
(a) starting point,
(b) optimized device, and 
(c) Luneburg lens approximation.}
\label{fig:opt_ra}
\end{figure}

\begin{figure}[htb]
\centering
\includegraphics{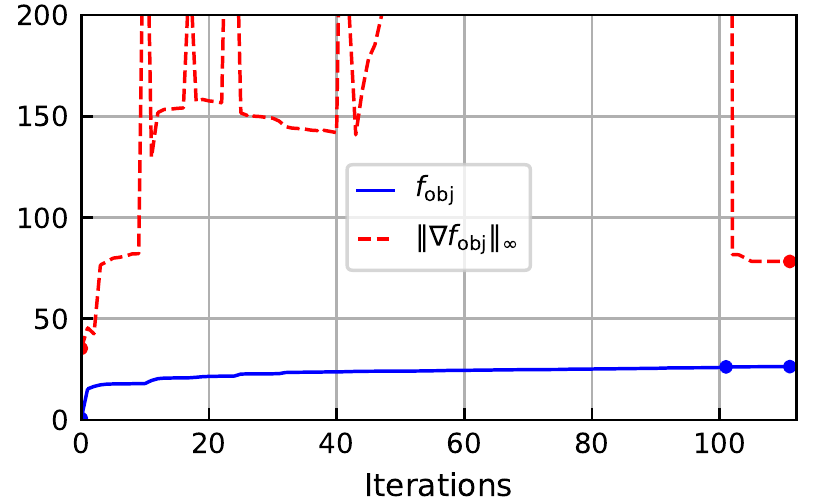}
\caption{Convergence progress of $f_\mathrm{obj}$ and its gradient norm for Fig.~\ref{fig:opt_ra} as a function of the penalized BFGS iteration. Markers indicate the beginning of an outer iteration. }
\label{fig:opt_rb}
\end{figure}

In Fig.~\ref{fig:opt_ra}, we consider focusing of an $\hat{\Vx}$-traveling plane wave to the focal point $(R_{\mathrm{lens}}, 0)$ on the lens rim.
In this example, there are 316 circular inclusions with relative permittivity $\varepsilon_r=4.5$, placed on a square grid with lattice constant $a = 0.2 \lambda_0$. The total lens radius is $R_{\mathrm{lens}} = 10a$, while the cylindrical harmonics parameter is $P=5$, and the initial guess is $R_m=a/4$ for all inclusions.
The penalized BFGS algorithm converged to a local maximum of $f_{\mathrm{obj}} = 26.36$ after $113$ total iterations and $173$ seconds, with the convergence criterion $\Delta R < 10^{-6}$, as shown in Fig.~\ref{fig:opt_rb}.
{Visualization 1} shows the electric field amplitude throughout the optimization process in video form, where the gradual evolution of the optimized device is clearly visible.

The optimization process yields a device that 
focuses the incoming electric field substantially better than the Luneburg lens, improving upon the Luneburg design by an amplitude factor of $1.55$. Additionally, the optimized design is more intricate than typical intuitive approximations, thus corroborating our promotion of an automated approach.
Interestingly, the algorithm produced symmetric radii with respect to the $x$ axis, although this was not an optimization constraint. Applying this constraint, thereby halving the optimization variables, yields a similar result in only $81$ seconds, less than half of the time required originally.
The optimized device is more susceptible than the Luneburg device to manufacturing variations, with a gradient norm of $78.3$, vs.~$68.4$ for the Luneburg. However, due to the significant improvement in performance we posit that the optimized device will outperform even with small radius perturbations.

\section{Conclusion}
\label{sec:conc}
We proposed and implemented an automated approach for designing dielectric metamaterials with desired electromagnetic properties.
Our approach uses gradient-based optimization that provides quick and reliable convergence as well as a fast boundary integral equation solver for precisely  computing the field at any point.
This method reduces the need for manual trial and error in the design of certain metamaterials by replacing it with rigorous optimization.
Our approach should be especially attractive in designing photonic crystals, metalenses, and other devices composed of many substructures whose large number of design parameters would typically render optimal manual design impossible.
Although optimization may superficially seem prohibitively expensive for these high-dimensional design problems, our fast solution method makes it practical. The examples in this paper resulted in highly irregular structures, which conforms to observations previously made in \cite{ar:seliger2006}, where the authors note that aperiodic structures are capable of providing more functionality than their periodic counterparts.
We implemented the methods described in this paper for the publicly available open-source software package {Particle}{Scattering.jl}~\cite{ar:blankrot2018joss} in the Julia programming language \cite{ar:julia2017review}, which also includes the examples presented here.

\section*{Funding}
Austrian Science Fund (FWF) START Project Y 660.

\bibliographystyle{ieeetran}
\bibliography{mybibfile}

\begin{thebibliography}{10}
\providecommand{\url}[1]{#1}
\csname url@samestyle\endcsname
\providecommand{\newblock}{\relax}
\providecommand{\bibinfo}[2]{#2}
\providecommand{\BIBentrySTDinterwordspacing}{\spaceskip=0pt\relax}
\providecommand{\BIBentryALTinterwordstretchfactor}{4}
\providecommand{\BIBentryALTinterwordspacing}{\spaceskip=\fontdimen2\font plus
\BIBentryALTinterwordstretchfactor\fontdimen3\font minus
  \fontdimen4\font\relax}
\providecommand{\BIBforeignlanguage}[2]{{%
\expandafter\ifx\csname l@#1\endcsname\relax
\typeout{** WARNING: IEEEtran.bst: No hyphenation pattern has been}%
\typeout{** loaded for the language `#1'. Using the pattern for}%
\typeout{** the default language instead.}%
\else
\language=\csname l@#1\endcsname
\fi
#2}}
\providecommand{\BIBdecl}{\relax}
\BIBdecl

\bibitem{ar:cloaking}
D.~Schurig, J.~J. Mock, B.~J. Justice, S.~A. Cummer, J.~B. Pendry, A.~F. Starr,
  and D.~R. Smith, ``Metamaterial electromagnetic cloak at microwave
  frequencies,'' \emph{Science}, vol. 314, no. 5801, pp. 977--980, 2006.

\bibitem{ar:pendry}
J.~B. Pendry, ``Negative refraction makes a perfect lens,'' \emph{Phys. Rev.
  Lett.}, vol.~85, pp. 3966--3969, Oct 2000.

\bibitem{ar:jahani}
S.~Jahani and Z.~Jacob, ``All-dielectric metamaterials,'' \emph{Nature
  Nanotechnology}, vol.~11, no.~1, pp. 23--36, 2016.

\bibitem{ar:yang2014}
Y.~Yang, W.~Wang, P.~Moitra, I.~I. Kravchenko, D.~P. Briggs, and J.~Valentine,
  ``Dielectric meta-reflectarray for broadband linear polarization conversion
  and optical vortex generation,'' \emph{Nano Letters}, vol.~14, no.~3, pp.
  1394--1399, 2014.

\bibitem{ar:moitra2015}
P.~Moitra, B.~A. Slovick, W.~Li, I.~I. Kravchencko, D.~P. Briggs,
  S.~Krishnamurthy, and J.~Valentine, ``Large-scale all-dielectric metamaterial
  perfect reflectors,'' \emph{ACS Photonics}, vol.~2, no.~6, pp. 692--698,
  2015.

\bibitem{ar:yablonovitch1994}
E.~Yablonovitch, ``Photonic crystals,'' \emph{Journal of Modern Optics},
  vol.~41, no.~2, pp. 173--194, 1994.

\bibitem{bk:joannopoulos2011}
J.~D. Joannopoulos, S.~G. Johnson, J.~N. Winn, and R.~D. Meade, \emph{Photonic
  Crystals: Molding the Flow of Light}.\hskip 1em plus 0.5em minus 0.4em\relax
  Princeton University Press, 2011.

\bibitem{ar:yablonovite}
E.~Yablonovitch, T.~J. Gmitter, and K.~M. Leung, ``Photonic band structure: The
  face-centered-cubic case employing nonspherical atoms,'' \emph{Phys. Rev.
  Lett.}, vol.~67, pp. 2295--2298, Oct 1991.

\bibitem{ar:turner2013}
M.~D. Turner, M.~Saba, Q.~Zhang, B.~P. Cumming, G.~E. Schr{\"o}der-Turk, and
  M.~Gu, ``Miniature chiral beamsplitter based on gyroid photonic crystals,''
  \emph{Nature Photonics}, vol.~7, no.~10, pp. 801--805, 2013.

\bibitem{ar:taverne2016}
M.~P.~C. Taverne, Y.-L.~D. Ho, X.~Zheng, S.~Liu, L.-F. Chen, M.~Lopez-Garcia,
  and J.~G. Rarity, ``Modelling defect cavities formed in inverse
  three-dimensional rod-connected diamond photonic crystals,'' \emph{EPL
  (Europhysics Letters)}, vol. 116, no.~6, p. 64007, 2016.

\bibitem{ar:yablonovitch2002photonic}
E.~Yablonovitch, ``Photonic bandgap based designs for nano-photonic integrated
  circuits,'' in \emph{International Electron Devices Meeting, 2002.
  IEDM'02}.\hskip 1em plus 0.5em minus 0.4em\relax IEEE, 2002, pp. 17--20.

\bibitem{ar:Cuesta-Soto:04}
F.~Cuesta-Soto, A.~Mart{\'i}nez, J.~Garc{\'i}a, F.~Ramos, P.~Sanchis,
  J.~Blasco, and J.~Mart{\'i}, ``All-optical switching structure based on a
  photonic crystal directional coupler,'' \emph{Opt. Express}, vol.~12, no.~1,
  pp. 161--167, Jan 2004.

\bibitem{ar:metalens_review}
M.~Khorasaninejad and F.~Capasso, ``Metalenses: Versatile multifunctional
  photonic components,'' \emph{Science}, 2017.

\bibitem{ar:khorasaninejad2016chiral}
M.~Khorasaninejad, W.~Chen, A.~Zhu, J.~Oh, R.~Devlin, D.~Rousso, and
  F.~Capasso, ``Multispectral chiral imaging with a metalens,'' \emph{Nano
  Letters}, vol.~16, no.~7, pp. 4595--4600, 2016.

\bibitem{ar:arbabi2016miniature}
A.~Arbabi, E.~Arbabi, S.~M. Kamali, Y.~Horie, S.~Han, and A.~Faraon,
  ``Miniature optical planar camera based on a wide-angle metasurface doublet
  corrected for monochromatic aberrations,'' \emph{Nature Communications},
  vol.~7, p. 13682, 2016.

\bibitem{ar:metalens_fiber}
N.~Yu and F.~Capasso, ``Optical metasurfaces and prospect of their applications
  including fiber optics,'' \emph{Journal of Lightwave Technology}, vol.~33,
  no.~12, pp. 2344--2358, June 2015.

\bibitem{ar:metasurfaces2014}
------, ``Flat optics with designer metasurfaces,'' \emph{Nature Materials},
  vol.~13, pp. 139--150, 2014.

\bibitem{ar:dobson1993}
D.~C. Dobson, ``Optimal design of periodic antireflective structures for the
  {Helmholtz} equation,'' \emph{European Journal of Applied Mathematics},
  vol.~4, no.~4, pp. 321–--339, 1993.

\bibitem{ar:bauer2008}
C.~A. Bauer, G.~R. Werner, and J.~R. Cary, ``Truncated photonic crystal
  cavities with optimized mode confinement,'' \emph{Journal of Applied
  Physics}, vol. 104, no.~5, p. 053107, 2008.

\bibitem{ar:cao2014}
Y.~Cao, J.~Xie, Y.~Liu, and Z.~Liu, ``Modeling and optimization of photonic
  crystal devices based on transformation optics method,'' \emph{Optics
  Express}, vol.~22, no.~3, pp. 2725--2734, 2014.

\bibitem{ar:seliger2006}
P.~Seliger, M.~Mahvash, C.~Wang, and A.~F.~J. Levi, ``Optimization of aperiodic
  dielectric structures,'' \emph{Journal of Applied Physics}, vol. 100, no.~3,
  p. 034310, 2006.

\bibitem{ar:bertsimas2007}
D.~Bertsimas, O.~Nohadani, and K.~M. Teo, ``Robust optimization in
  electromagnetic scattering problems,'' \emph{Journal of Applied Physics},
  vol. 101, no.~7, p. 074507, 2007.

\bibitem{ar:bao1998modeling}
G.~Bao and D.~C. Dobson, ``Modeling and optimal design of diffractive optical
  structures,'' \emph{Surveys on Mathematics for Industry}, vol.~8, no.~1, pp.
  37--62, 1998.

\bibitem{ar:yablonovitch2005}
C.~Y. Kao, S.~Osher, and E.~Yablonovitch, ``Maximizing band gaps in
  two-dimensional photonic crystals by using level set methods,'' \emph{Applied
  Physics B}, vol.~81, no.~2, pp. 235--244, July 2005.

\bibitem{ar:jesselu2013}
J.~Lu and J.~Vu\v{c}kovi\'{c}, ``Nanophotonic computational design,''
  \emph{Optics Express}, vol.~21, no.~11, pp. 13\,351--13\,367, Jun 2013.

\bibitem{ar:miller2013adjoint}
C.~Lalau-Keraly, S.~Bhargava, O.~D. Miller, and E.~Yablonovitch, ``Adjoint
  shape optimization applied to electromagnetic design,'' \emph{Optics
  Express}, vol.~21, pp. 21\,693--21\,701, 2013.

\bibitem{ar:gumerov2007}
N.~A. Gumerov and R.~Duraiswami, ``A scalar potential formulation and
  translation theory for the time-harmonic maxwell equations,'' \emph{Journal
  of Computational Physics}, vol. 225, no.~1, pp. 206--236, 2007.

\bibitem{ar:lai14}
J.~Lai, M.~Kobayashi, and L.~Greengard, ``A fast solver for multi-particle
  scattering in a layered medium,'' \emph{Opt. Express}, vol.~22, no.~17, pp.
  20\,481--20\,499, Aug 2014.

\bibitem{bk:kress2013}
D.~Colton and R.~Kress, \emph{Integral Equation Methods in Scattering
  Theory}.\hskip 1em plus 0.5em minus 0.4em\relax New York: Wiley, 1983.

\bibitem{ar:lax1951}
M.~Lax, ``Multiple scattering of waves,'' \emph{Rev. Mod. Phys.}, vol.~23, pp.
  287--310, Oct 1951.

\bibitem{ar:coifman1993}
R.~Coifman, V.~Rokhlin, and S.~Wandzura, ``The fast multipole method for the
  wave equation: a pedestrian prescription,'' \emph{IEEE Antennas and
  Propagation Magazine}, vol.~35, no.~3, pp. 7--12, June 1993.

\bibitem{ar:chew1991}
L.~Gurel and W.~C. Chew, ``On the connection of {T} matrices and integral
  equations,'' in \emph{Antennas and Propagation Society Symposium 1991
  Digest}, vol.~3, June 1991, pp. 1624--1627.

\bibitem{ar:martin2003}
P.~A. Martin, ``On connections between boundary integral equations and
  {T}-matrix methods,'' \emph{Engineering Analysis with Boundary Elements},
  vol.~27, no.~7, pp. 771--777, 2003.

\bibitem{ar:greengard2013}
Z.~Gimbutas and L.~Greengard, ``Fast multi-particle scattering: A hybrid solver
  for the {Maxwell} equations in microstructured materials,'' \emph{Journal of
  Computational Physics}, vol. 232, no.~1, pp. 22--32, 2013.

\bibitem{ar:chavent1974identification}
G.~Chavent, ``Identification of functional parameters in partial differential
  equations,'' in \emph{Joint Automatic Control Conference}, no.~12, 1974, pp.
  155--156.

\bibitem{ar:plessix2006adjoint}
R.-E. Plessix, ``A review of the adjoint-state method for computing the
  gradient of a functional with geophysical applications,'' \emph{Geophysical
  Journal International}, vol. 167, no.~2, pp. 495--503, 2006.

\bibitem{ar:Ganesh2012}
M.~Ganesh, S.~C. Hawkins, and R.~Hiptmair, ``Convergence analysis with
  parameter estimates for a reduced basis acoustic scattering {T-matrix}
  method,'' \emph{IMA Journal of Numerical Analysis}, vol.~32, no.~4, pp.
  1348--1374, 2012.

\bibitem{ar:rokhlin83}
V.~Rokhlin, ``Solution of acoustic scattering problems by means of second kind
  integral equations,'' \emph{Wave Motion}, vol.~5, no.~3, pp. 257--272, 1983.

\bibitem{ar:bremer2010}
J.~Bremer, V.~Rokhlin, and I.~Sammis, ``Universal quadratures for boundary
  integral equations on two-dimensional domains with corners,'' \emph{Journal
  of Computational Physics}, vol. 229, no.~22, pp. 8259--8280, 2010.

\bibitem{ar:kress94}
R.~Kress, ``On the numerical solution of a hypersingular integral equation in
  scattering theory,'' \emph{Journal of Computational and Applied Mathematics},
  vol.~61, no.~3, pp. 345 -- 360, 1995.

\bibitem{ar:qbx13}
A.~Kl{\"o}ckner, A.~Barnett, L.~Greengard, and M.~O'Neil, ``Quadrature by
  expansion: A new method for the evaluation of layer potentials,''
  \emph{Journal of Computational Physics}, vol. 252, no. Supplement C, pp.
  332--349, 2013.

\bibitem{bk:chew2001fast}
W.~C. Chew, E.~Michielssen, J.~M. Song, and J.~M. Jin, \emph{Fast and Efficient
  Algorithms in Computational Electromagnetics}.\hskip 1em plus 0.5em minus
  0.4em\relax Norwood, MA, USA: Artech House, Inc., 2001.

\bibitem{ar:saad1986gmres}
Y.~Saad and M.~H. Schultz, ``{GMRES}: A generalized minimal residual algorithm
  for solving nonsymmetric linear systems,'' \emph{SIAM Journal on Scientific
  and Statistical Computing}, vol.~7, no.~3, pp. 856--869, 1986.

\bibitem{ar:julia2017review}
J.~Bezanson, A.~Edelman, S.~Karpinski, and V.~B. Shah, ``Julia: A fresh
  approach to numerical computing,'' \emph{SIAM Review}, vol.~59, no.~1, pp.
  65--98, 2017.

\bibitem{bk:optimization}
J.~Nocedal and S.~Wright, \emph{Numerical Optimization}, 2nd~ed.\hskip 1em plus
  0.5em minus 0.4em\relax Springer-Verlag New York, 2006.

\bibitem{ar:armijo}
L.~Armijo, ``Minimization of functions having {Lipschitz} continuous first
  partial derivatives,'' \emph{Pacific Journal of Mathematics}, vol.~16, no.~1,
  pp. 1--3, 1966.

\bibitem{bk:optics}
M.~Born and E.~Wolf, \emph{Principles of Optics: Electromagnetic Theory of
  Propagation, Interference and Diffraction of Light}, 6th~ed.\hskip 1em plus
  0.5em minus 0.4em\relax New York: Pergamon Press, 1980.

\bibitem{ar:luneburg2010}
S.~Takahashi, C.-H. Chang, S.~Y. Yang, and G.~Barbastathis, ``Design and
  fabrication of dielectric nanostructured {L}uneburg lens in optical
  frequencies,'' in \emph{2010 International Conference on Optical MEMS and
  Nanophotonics}, Aug 2010, pp. 179--180.

\bibitem{ar:blankrot2018joss}
B.~Blankrot and C.~Heitzinger, ``{ParticleScattering}: Solving and optimizing
  multiple-scattering problems in {Julia},'' \emph{Journal of Open Source
  Software}, vol.~3, no.~25, p. 691, May 2018.

\end{thebibliography}

\end{document}